\documentclass[doc]{article}

\usepackage[margin=1in]{geometry}
\usepackage{graphicx, xcolor, geometry}
\usepackage{booktabs}
\usepackage{tabularx}
\usepackage{amsmath}
\usepackage{bm}
\usepackage{epstopdf}
\usepackage{amssymb}
\usepackage{multirow}
\usepackage{url}
\usepackage{color} 
\usepackage{soul}
\usepackage[textsize=tiny]{todonotes}
\usepackage[title]{appendix}
\usepackage{titlesec}
\usepackage{authblk}

\newcolumntype{L}{>{\arraybackslash}m{4cm}}
\newcolumntype{Q}{>{\arraybackslash}m{10cm}}

\titleformat{\paragraph}
{\normalfont\normalsize\bfseries}{\theparagraph}{1em}{}
\titlespacing*{\paragraph}
{0pt}{3.25ex plus 1ex minus .2ex}{1.5ex plus .2ex}

\graphicspath{{figs/}}
\usepackage{caption}

\makeatletter
\renewcommand*\env@matrix[1][\arraystretch]{%
 \edef\arraystretch{#1}%
 \hskip -\arraycolsep
 \let\@ifnextchar\new@ifnextchar
 \array{*\c@MaxMatrixCols c}}
\makeatother

\begin{document}

\title{Experimental Design Modulates Variance in BOLD Activation: The Variance Design General Linear Model}


\author[1]{Garren Gaut}
\author[2,3]{Xiangrui Li}
\author[2,3]{Zhong-Lin Lu }
\author[1]{Mark Steyvers  }

\affil[1]{Department of Cognitive Science, University of California Irvine, Irvine CA, USA}
\affil[2]{Center for Cognitive and Behavioral Brain Imaging, The Ohio State University, Columbus OH, USA}
\affil[3]{Department of Psychology, The Ohio State University, Columbus OH, USA}
\date{} 

\maketitle

\begin{abstract}

Typical fMRI studies have focused on either the mean trend in the blood-oxygen-level-dependent (BOLD) time course or functional connectivity (FC). However, other statistics of the neuroimaging data may contain important information. Despite studies showing links between the variance in the BOLD time series (BV) and age and cognitive performance, a formal framework for testing these effects has not yet been developed. 

We introduce the Variance Design General Linear Model (VDGLM), a novel framework that facilitates the detection of variance effects. We designed the framework for general use in any fMRI study by modeling both mean and variance in BOLD activation as a function of experimental design. The flexibility of this approach allows the VDGLM to i) simultaneously make inferences about a mean or variance effect while controlling for the other and ii) test for variance effects that could be associated with multiple conditions and/or noise regressors. We demonstrate the use of the VDGLM in a working memory application and show that engagement in a working memory task is associated with whole-brain decreases in BOLD variance. 

\end{abstract}


\section{ Introduction }

At their core, neuroimaging analyses consist of relating a summary statistic of the  blood-oxygen-level-dependent (BOLD) time course to experimental condition, behavior, or individual characteristics. The primary method for fMRI analysis, the General Linear Model (GLM) \cite{friston1994analysis, bullmore1996statistical}, focuses on the mean trend in the BOLD activation. Recently, researchers have moved beyond mean BOLD trend by studying functional connectivity (FC), which is calculated as the Pearson correlation between regions. However, it is possible that other statistics of the neuroimaging data may contain important information. A natural candidate is BOLD variability (BV) defined as the variance in the BOLD time series. BV can be thought of as intermediate to mean BOLD trend and FC in terms of computational complexity; BV is based on a locally independent computations whereas FC incorporates between-region dependencies. Importantly, as FC and mean BOLD trend have led to distinct avenues of research, BV could be a fundamentally different channel for studying brain function. 

This article introduces the Variance Design General Linear Model (VDGLM), a novel framework that allows researchers to simultaneously test for effects on BV and on mean BOLD activation. The VDGLM can be conceptualized as a GLM that explicitly incorporates the experimental design into the model of the variance. Direct incorporation of the experimental design allows the VDGLM to be flexible enough to be used in any fMRI experiment. This new framework facilitates the analysis of BV effects and enables new discoveries that relate BV to disease, individual characteristics, and human behavior. 

The development of the VDGLM was motivated in part by studies in EEG and fMRI that have demonstrated relationships between brain fluctuations and cognitive processes, behavior, and age. EEG studies have shown variance effects in the form of suppression of alpha and theta oscillatory waves, that is, reduction of the amplitude of the oscillations (see \cite{ klimesch1999eeg} for a review). Alpha suppression is related to task engagement \cite{williamson1997study}, opening the eyes \cite{berger1929elektrenkephalogramm}, sleep \cite{dement1957cyclic}, and cognitive performance \cite{klimesch1999eeg}. Age studies have found that alpha, delta, and theta  suppression increases with age during youth \cite{somsen1997growth} and that alpha suppression decreases with age in older populations \cite{duffy1984age}. Alpha suppression's relationship to cognitive and individual differences are mediated by inhibitory-control processes involved in attention \cite{klimesch2005functional,klimesch2012alpha}.

In fMRI, brain fluctuations measured by BV have been shown to vary with behavior and age. BV varies between task and fixation, particularly in younger adults \cite{garrett2012modulation}. Differences in BV between task and fixation are associated with higher visual discrimination performance \cite{wutte2011physiological} and track task difficulty \cite{garrett2013brain}. Age has also been related to BV; BV was shown to predict age with five times the explanatory power of mean BOLD \cite{garrett2010blood} and was indicative of younger, faster, and more consistently performing subjects \cite{garrett2011importance}. In both studies, the spatial distribution of BV effects was orthogonal to the distribution of mean effects. Furthermore, a follow-up study found that age related BV affects were robust to vascular controls \cite{garrett2017age}. Gaut et al. found that BV could be used to accurately predict the task a subject was performing and subject identity \cite{gaut2018predicting}. Despite these links to behavior and age, the study of BOLD variability has not been widely pursued in fMRI. 

Whereas the aforementioned studies focused on BV, the studies did not use a general statistical framework for expressly studying BV effects. Therefore, another key motivation for the VDGLM framework is to introduce a unified fMRI framework to allow for the analysis of BV. The framework that we develop, the VDGLM, is a parametric approach that jointly models the mean and variance by explicitly incorporating the experimental design into the variance formulation. The inclusion of the design in the variance allows us to: {\it i)} jointly model mean and variance effects, { \it ii)} explicitly model the temporal dynamics between BV and experimental condition, and { \it iii)} include multiple experimental conditions in our variance analyses. The explicit structure of mean and variance design is supplied by the researcher, which allows for easy generalization to any experiment, and model fitting is computationally efficient enough to run region of interest (ROI) analyses on large brain imaging studies. By developing this framework, we are providing an important tool to test for variance effects that has the potential to spur new research developments in various fields. 



The plan for the rest of the paper is as follows. We first provide an overview of the GLM, and establish the theory behind the VDGLM. Next we provide an application of the VDGLM to Working Memory data from the Human Connectome Project Healthy Adult dataset. We finish with a discussion of the choices made when using the VDGLM and how variance measured by the VDGLM compares to other measures of variance. 

\begin{figure}[!h]
  \centering 
  	\includegraphics[width=0.49\textwidth]{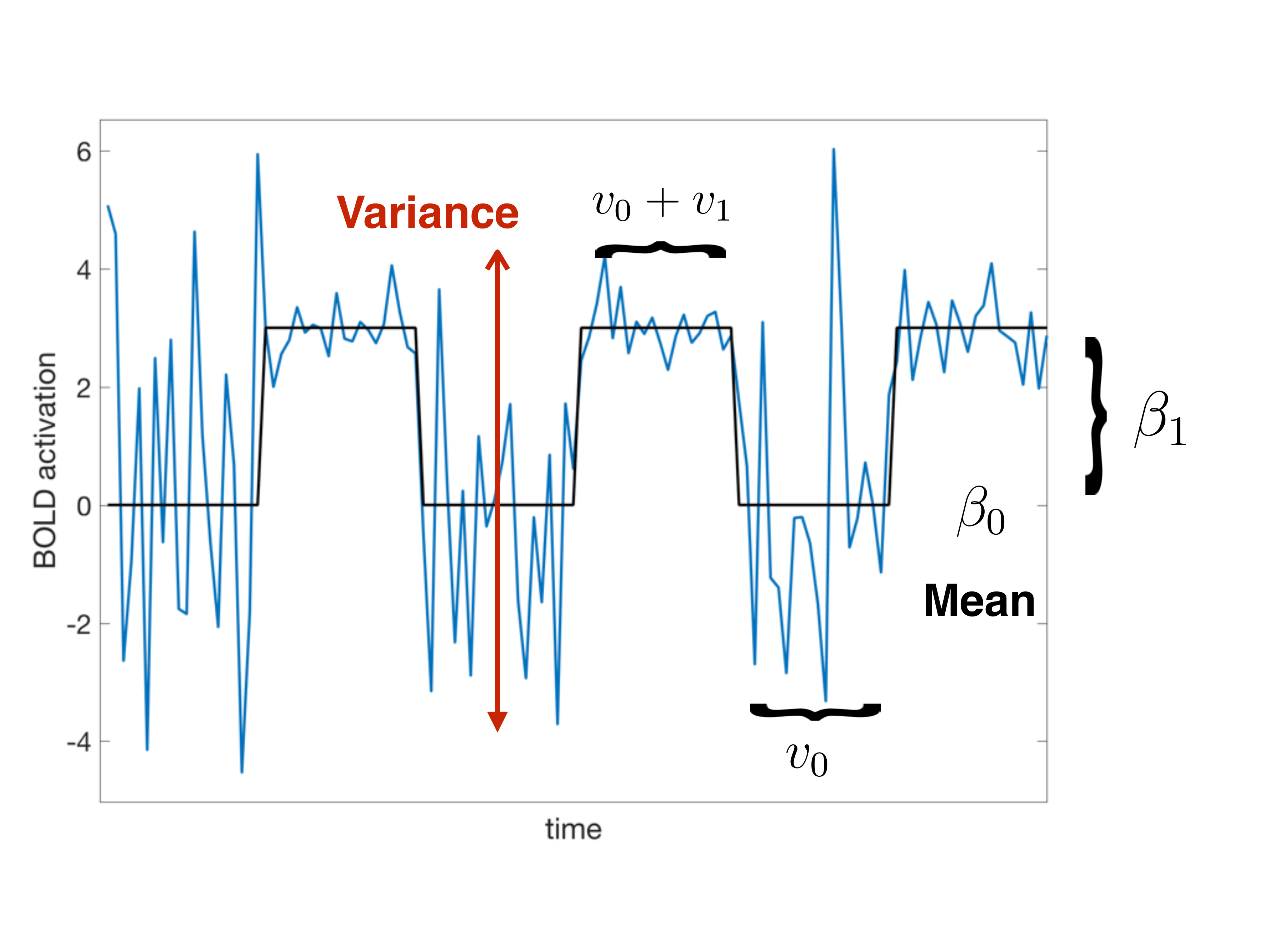}
  	\caption{Illustration of artificial data where the presence of a single experimental condition (black) increases the mean but lowers the variance of the BOLD time course (blue). The values used to create this visualization are based on Equation \ref{simpleVDGLM} with $\beta_0 = 0, \beta_1 = 3, v_0 = 2,$ and $v_1 = -1.5$}
  	\label{mean_var_example}
\end{figure}

\section{ A Novel Framework for Studying BV } 

To motivate the VDGLM, consider a hypothetical BOLD activation time series where BV is affected by an experimental condition that indicates fixation versus task (see Figure \ref{mean_var_example}). The single experimental condition is plotted in black and the BOLD time series from a single voxel is plotted in blue. For conceptual simplicity, the experimental condition time series is not convolved with a hemodynamic response function (HRF) model. The voxel time series varies as a function of task condition; the variance is higher during fixation compared to task. We can describe these effects on the mean and variance using a simple VDGLM:

\begin{equation}\label{simpleVDGLM}
y \sim N(\beta_0 + x^T\beta_1 , (v_0 + x^T v_1)I)
\end{equation}

\noindent where $y$ is the BOLD time series, $x$ represents the condition indicator, $I$ is the identity matrix, $\beta_0$ captures the mean activation, and $v_0$ captures the measurement variance, i.e., the out-of-task variation. Then $\beta_1$ and $v_1$ capture the degree of change in mean and variance due to task engagement, respectively. If the model is applied to the data from Figure \ref{mean_var_example}, we expect $\beta_0 \approx 0$, $\beta_1 \approx 3$, $v_0$ to be a large positive value, and $v_1$ to be negative, but with the constraint that $|x_tv_1| < |v_0| \; \forall t$. Here, the parameter $v_1 < 0$ reflects the fact that BOLD variation is lower within task compared to fixation. 

For comparison, the GLM estimates a single variance parameter over the entire time series and ignores the change in variance due to the experimental manipulation: 

\begin{equation}\label{simpleGLM}
y \sim N(\beta_0 + x^T\beta_1 , v_0I)
\end{equation}

\noindent Note that Eq. \ref{simpleGLM} is equivalent to Eq. \ref{simpleVDGLM} when $v_1 = 0$, i.e., the GLM is a nested model of the VDGLM where there is no experimental modification of the variance. The GLM is a null model for no effect of the variance that can be compared to the VDGLM {\it i}) to explicitly test for variance inclusion and {\it ii)} test whether the mean effects found by the VDGLM are similar to the mean effects found by the GLM.

\subsection{ VDGLM Analysis Pipeline }

\begin{figure*}[!h]
  \centering 
 	\includegraphics[width=1\textwidth]{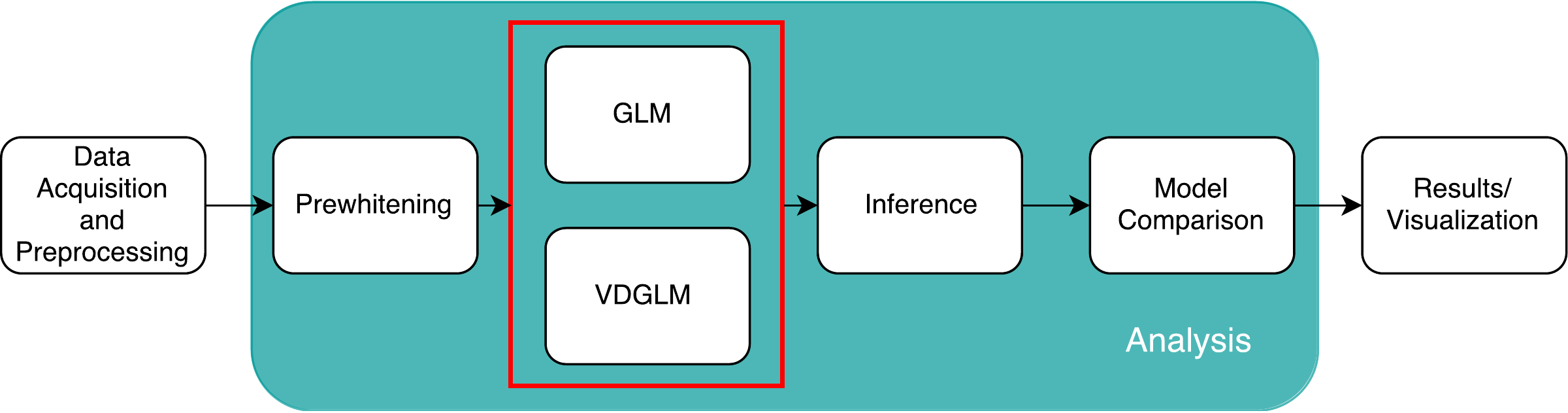}
 	\caption{ An illustration of a typical fMRI pipeline that uses either the GLM or the VDGLM for analysis. To use the VDGLM, the only steps from a traditional pipeline that must change are model formulation and estimation. Data acquisition, preprocessing, prewhitening, model comparison, and methods for result dissemination can remain the same. Some inference steps, such as effect size estimation, could remain the same. However, other inference procedures, such as significance testing, require statistics developed expressely for the VDGLM. }
 	\label{fmri_flow}
\end{figure*}

One goal of the VDGLM framework is to allow the VDGLM to be inserted into any standard fMRI analysis pipeline with minimal modifications (see Figure \ref{fmri_flow}). The VDGLM does not change data acquisition, preprocessing, prewhitening, model comparison, or results dissemination. The main step that must change is model formulation and estimation. In some cases, the inference step is not affected (e.g., computing effect sizes using Cohen's d). However, more sophisticated inference such as parameter significance testing will require modification of the inference step to include statistics for testing variance parameters (see section \ref{group_level_analysis}). In a BV focused analysis, we also recommend additional preprocessing steps to remove variance confounds such as censoring and head motion correction, but these additional steps are not necessary to use the VDGLM.

\subsection{ Matrix Notation }



We can write the GLM and the VDGLM in matrix notation to highlight the concept of inserting the design matrix into the variance formulation. The GLM models the BOLD time series $y_{[T\times 1]}$ from a single voxel as a linear function of the experimental design \cite{friston1994statistical, bullmore1996statistical, beckmann2003general, woolrich2004multilevel}. Formally, the GLM is defined: 

\begin{align}\label{GLM} 
y &= X \beta + \epsilon \\ 
\epsilon & \sim N(0, I\sigma^2) \nonumber
\end{align} 

\noindent where $X$ is a $T \times p$ design matrix, $\beta$ is a $ p \times 1$ vector of mean parameters, $\sigma^2$ is a variance parameter, and $I$ is the identity matrix. The columns of the design matrix $X$ include experimental events, experimental blocks, stimulus presentation, or mean activation. The VDGLM has the same formulation, but extends the variance model: 

\begin{align} \label{VDGLM}
y &= X_m\beta + \eta \\ 
\eta & \sim N(0, \textrm{diag}(X_v v)I)  \nonumber \\ 
& \textrm{diag}(X_v v)I > 0 \nonumber
\end{align} 

\noindent where diag$(x)$ is the matrix with the entries of the vector $x$ along the diagonal. To emphasize that the mean and variance design matrices can be distinct, we use the notation $X_m$ and $X_v$ to denote the mean and variance designs, respectively. The parameters, $\beta$ and $v$, capture mean and variance effects, respectively. It is clear that the GLM (eq. \ref{GLM}) is a special case of the VDGLM for which the variance design matrix $X_v$ is a single column of ones and $v = \sigma^2$. 

\subsubsection{ Prewhitening and Noise Regressors} 

In GLM analyses, BOLD time serires are 'prewhitened' to account for BOLD autocorrelation  \cite{bullmore1996statistical, woolrich2001temporal}. We also prewhiten before fitting the VDGLM to ensure that variance effects found by the VDGLM are not artifacts caused by autocorrelation. In VDGLM, as with GLM analyses, any standard autocorrelation estimator can be used \cite{woolrich2001temporal, friston2000smooth, cox1996afni}. In theory, one could use the residuals from either the VDGLM or the GLM to estimate the autocorrelation. We recommend using the GLM residuals for two reasons. First, the GLM is less computationally intensive than the VDGLM, and we expect that the VDGLM leads to similar residuals since in practice we've found the mean trend for the VDGLM to be similar to the mean trend for the GLM when fit to unwhitened data. Second, by using GLM residuals, we decouple variance effects due to autocorrelation and true variance effects, i.e., any variance signal that could be accounted for by either autocorrelation or the VDGLM is by default attributed to autocorrelation. By removing this autocorrelation using prewhitening, we ensure that the whitened data will lead to more conservative estimation of variance effects than if we had used the VDGLM for prewhitening (i.e., it is less likely that artifactual autocorrelation will lead to detection of variance effects). 

Other techniques for controlling noise  (e.g., coloring or head motion correction)  involve the addition of noise regressors. As with GLM analyses, the VDGLM can incorporate these techniques by including the appropriate regressors in the design matrices. 

\subsubsection{ Estimation } \label{param_estimation}
Univariate GLM mean parameter estimation can proceed in one of two ways: {\it i}) ordinary or general least squares \cite{friston1994statistical, beckmann2003general} or {\it ii}) fully Bayesian inference \cite{woolrich2004multilevel}. Approximate Bayesian inference has also been used, but in a single group-level analysis that combines first and second-level models \cite{friston2002classical}. Variance estimation is usually done using by iteratively computing OLS estimates \cite{woolrich2001temporal, worsley2002general}, but can require more advanced methods depending on the structure of group-level models (see section \ref{group_level_analysis}).   

There are potentially many approaches that could be used to estimate the VDGLM, including Bayesian and maximum likelihood approaches. For simplicity, we use a maximum likelihood approach. Estimation approaches must be computationally efficient enough to handle the high dimensionality of fMRI data and in practice, we found that sampling techniques were too slow to be practical for large data sets. Maximum likelihood (or maximum a posteriori) estimation using mode-finding algorithms is efficient enough to estimate parameters for an ROI analysis from a large fMRI dataset in about half a day using parallel computing techniques. 

There is known bias in variance estimates when computing MLE solutions \cite{harville1977maximum}. However, in fMRI studies, the large number of time measurements will lead to small biases that can be considered negligible  (e.g., a bias of 1/405 $\approx 0.0025$ for the Working Memory Task of the HCP Healthy Adults data) .  


\subsubsection{ Group Level Analysis } \label{group_level_analysis}

Group-level GLM analyses typically incorporate two-stages, in which second stage analysis is based on summary statistics from the first \cite{holmes1998generalisability,beckmann2003general,woolrich2004multilevel}. The methodology for group-level significance testing depends on the experimental design. T-tests can be used provided that the experiment is balanced \cite{holmes1998generalisability}. For unbalanced data, if the variance components of the data are known, then principled group-level inference can be done using univariate parameter estimates and their covariance estimates \cite{beckmann2003general}. In most cases, these variance components are not known. Second-level variance parameter estimates have been found using the EM algorithm \cite{worsley2002general}, approximate Bayesian inference \cite{friston2002classical} and fully Bayesian inference \cite{woolrich2004multilevel}. These same ideas extend to the VDGLM. For the simple balanced-experiment setting, group-level inference can be computed using t-tests. In the unbalanced case, more work is needed due to the difficulty in computing the covariance of variance parameters. We initially tried to develop group-level inference procedures using asymptotic statistics (Wald test), but these tests were ill-behaved for several subjects due to high-condition number matrix inversions. We leave development of alternative statistics and a fully Bayesian framework to future work.

Group effect sizes can be estimated using the set of parameter estimates from all subjects. In our application, we compute Cohen's d, which measures the standardized mean between two populations, and is popular in fMRI for avoiding the multiple comparison problem inherent in significance testing.


\subsubsection{ Model Comparison } 

Model comparison also proceeds as in a traditional fMRI pipeline. Model comparison can be done using AIC \cite{akaike1974new}, BIC \cite{schwarz1978estimating}, or any other log-likelihood-based metric that is a function of a point estimate. Model comparisons can consist of traditional in-sample comparisons or can be generalized to new data using out-of-sample comparisons \cite{mosteller1968data}. The outcomes of univariate comparisons can be aggregated into group level results that test whether a subject tends to prefer a certain model across the brain or whether a particular region tends to prefer a certain model across subjects.

\section{ Example Application: BV in Working Memory }

In this example application, we used the VDGLM to find brain regions that are involved in working memory via changes in BV. We examined whether these regions differ from regions involved via changes in mean BOLD activation, and tested whether the VDGLM better describes working memory data than the GLM. The goal is to illustrate how to use the VDGLM and to showcase its utility. 

We used data from the Human Connectome Project (HCP) Working Memory Experiment. In the experiment, subjects alternated between fixation blocks and two different task blocks during which they were presented with sequences of visual stimuli. In a 2-back task block, subjects indicated whether the current stimulus was the same as the one two presentations ago. In a 0-back task block, subjects indicated when a target stimulus was presented. 

GLM analyses of the HCP data have found that engagement in the 2-back working memory task invokes regions thought to be involved in a cognitive control network, including bilateral dorsal and ventral prefrontal cortex, dorsal parietal cortex and dorsal anterior cingulate. Task engagement leads to a deactivation in the default mode network, namely in the medial prefrontal cortex, posterior cingulate, and the occipital parietal junction \cite{barch2013function}. Similar activation patterns are found even when comparing 2-back versus 0-back. A 24 study meta-analysis of N-back studies found consistent activation in frontal and parietal areas, namely bilateral and
medial posterior parietal cortex, bilateral
premotor cortex, dorsal cingulate/medial premotor
cortex, bilateral rostral prefrontal cortex or frontal poles, bilateral dorsolateral prefrontal cortex, and bilateral mid-ventrolateral prefrontal cortex or frontal
operculum \cite{owen2005n}. 

In our VDGLM analysis, the goal was to find both mean effects that overlap with known mean effects and also variance effects that could be spatially orthogonal to known mean effects.

\subsection{ Methods }

\subsubsection{ Data Acquisition and Preprocessing }

The data was collected by the Washington University - University of Minnesota Consortium Human Connectome Project (HCP, Van Essen et al., 2013). We used the Working Memory task data from the 1200 Subjects release using the minimal pre-processing pipeline \cite{glasser2013minimal}. Details of task fMRI processing can be found in \cite{barch2013function}. We included subjects that performed the left-to-right phase encoded Working Memory task, resulting in 875/1200 total subjects for analysis. The downloaded data were in grayornidate system \cite{glasser2013minimal}, and the time series for 333 surface regions of interest (ROIs) based on Gordon et al. were extracted for further analysis \cite{gordon2014generation}. We perform additional preprocessing including scrubbing and regression of motion estimates to minimize motion artifacts \cite{burgess2016evaluation}. The preprocessed version of our data is uploaded on the Open Science Foundation website: (https://osf.io/4rvbz/ \cite{gautOSF}). 

\subsubsection{ Task Design } 

\begin{figure}[!h]
  \centering 
 	\includegraphics[width=0.49\textwidth]{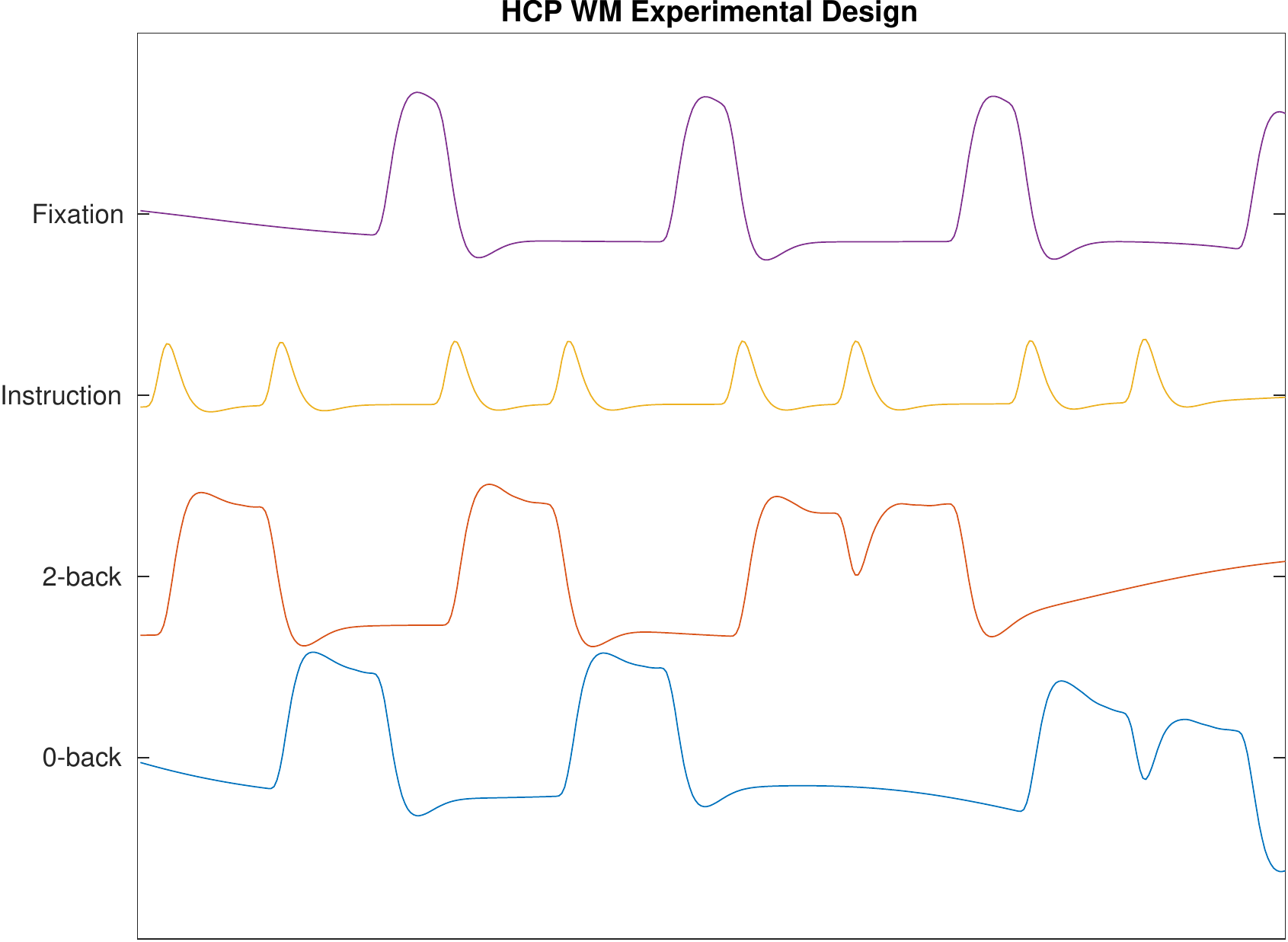}
 	\caption{HCP working memory experimental design convolved with the canonical HRF. Condition is indicated on the y-axis. }
 	\label{design}
\end{figure}

During the Working Memory experiment, subjects alternatively engaged in a 0-back and 2-back tasks that use faces, places, tools and body parts as the four categories of stimuli. Within each run, subjects were presented with blocks of stimuli, where all stimuli within a block were from the same category. For half of the blocks, subjects were given a ``target'' stimulus and were instructed to press a button whenever that stimulus was presented (0-back task). For the other half of blocks, subjects were instructed to respond when the stimulus was the same as the one presented two presentations ago (2-back task). Task blocks were interwoven with 15 second fixation blocks and instruction cues indicating the task type and 'target' stimulus if the task was the 0-back task. Each run contained 8 task blocks. We combined blocks from each stimuli type to create two task indicators (one for 0-back and one for 2-back). In total, the experimental design contained four conditions: the 0-back task, the 2-back task, Fixation, and Instruction (see Figure \ref{design}). 

\subsubsection{ Modeling } 

We applied the VDGLM model (eq. \ref{VDGLM}) to the data. Building the VDGLM required specifying both the mean and variance design matrix. In the mean design, we included the 0-back, 2-back, Fixation, and Instruction conditions. In the variance design we included the the same regressors as in the mean design, but with an additional intercept regressor to reflect the assumption that there exists some measurement noise not captured by the other variance regressors.

We also fit a GLM (eq. \ref{GLM}) model to the data using a design matrix that is equivalent to the mean design matrix used in the VDGLM.   

\paragraph{Prewhitening} 

The first step in model estimation is to prewhiten the data. We fit a GLM model from which we computed the residuals and then used an AR(2) process to estimate residual autocorrelation. We chose the AR(2) process because it has been shown to outperform standard autocorrelation estimators on tests of autocorrelation present after prewhitening \cite{lenoski2008performance}. An AR(2) process models the GLM residuals $r_t$ at time $t$ as: 
\begin{equation} 
r_t = \phi_1r_{t-1} + \phi_2r_{t-2} + \epsilon
\end{equation} 

\noindent where $\phi_i$ measure the contribution of the $i$-th autoregressive component and $\epsilon ~ N(0,\sigma^2)$ is white noise. We estimated the autoregressive parameters using the Yule-Walker equation \cite{yule1927vii,walker1931periodicity}, from which the estimated autocovariance was generated using a simple parametric form \cite{lenoski2008performance}. 

\paragraph{Estimation}

After prewhitening, we estimated GLM parameters using ordinary least squares. Since the VDGLM is analytically intractable, we estimated parameters using constrained trust-region optimization \cite{more1983computing} (see Appendix \ref{optimization} for optimization details, and \cite{yuan2000review} for a review of trust-region optimization). We performed mass univariate estimation, i.e., we fit the VDGLM and the GLM for each ROI and subject. From parameter estimates, we created parameter contrasts for the 2-back minus Fixation, 0-back minus Fixation, and 2-back minus 0-back conditions.

\paragraph{ Group Level Effect Sizes } 

We estimated group-level effect sizes for each contrast using Cohen's d (the difference in standardized means) computed over subjects \cite{cohen1977statistical}. 
We visualized these effect sizes for each ROI using the HCP workbench software \cite{marcus2011informatics}. For a single region there exist 3 possible group-level effect patterns on BOLD: 1) both mean and variance effects are shown, 2) either mean or variance effects are shown, or 3) neither type of effect is shown. We plotted the whole-brain spatial distribution of each type of effect at small and medium effect sizes (Cohen's d of 0.2, 0.5, respectively). Additionally, we compare our VDGLM mean estimates to GLM mean estimates, to see whether modeling the variance changes known mean inferences.


\paragraph{ Model Comparison } 

Because the VDGLM has more parameters than the GLM, it has the potential to explain more variability of the observed data, thus any model comparison metric should take complexity into account. We achieved this using out-of-sample log likelihood (OOSLL), which penalizes overfitting by testing how well a model generalizes to new unseen data. We used an out-of-sample metric, rather than traditional metrics of model fit (such as goodness of fit tests, or information criteria) to balance the goals of our analysis between prediction and explanation \cite{yarkoni2016choosing}; the model was compared to other models using predictive performance, but also evaluated on its explanatory power. We used 10-fold cross validation to compute the out-of-sample log likelihood. For a single subject and region, we split the time series into 10 folds that each contains a test and training set. For each fold, we fit our model on the training set and computed the out-of-sample log likelihood of the test set given the parameters computed during training. To understand the level of general preference for the VDGLM, we compute the percent of subject/ROI time series with higher OOSLL for the VDGLM compared to the GLM. To understand subject VDGLM preference, we compute the percent of regions that prefer the VDGLM model for each subject. 

To test that our model comparison results did not occur by chance, we compared the prevalence of VDGLM preference found in the real data to that found for a dataset simulated from the GLM (i.e., data without variance effects). In this comparison, we wanted to make explicit assurances that the VDGLM was preferred because of real effects in the data, i.e., that preference was not due to autocorrelation artifacts. We did this by adding autocorrelation to the dataset simulated from the GLM, where we estimated the autocorrelation from the real data. We generated a time series for each subject and ROI independently. The generation of a sample time series from a real time series $y_k$ proceeded as follows: 

\begin{enumerate} 
\item Compute the GLM OLS solution $\hat{\beta}$ and variance solution $\hat{\sigma}^2 $. We use GLM parameters estimated from the real data to better account for subject heterogeneity than if we simulated the underlying GLM parameters.
\item Estimate the autocorrelation of the residuals $y_k - X\hat{\beta}$ using an AR(2) process and generate an estimated autocovariance matrix $A$. 
\item Generate a sample time series $y_{\textrm{samp}} = X \hat{\beta} + \epsilon_{\textrm{samp}}$, where $\epsilon_{\textrm{samp}} \sim MVN( 0 , \hat{\sigma}^2A) $.
\end{enumerate} 


 
For each subject and region, we generate a single sample time series (we generate only one sample to reduce the computational complexity of this test). Using the generated dataset, we fit the VDGLM to each subject and region. Due to having just one sample from each subject and ROI, we cannot make statements about whether model comparisons for a single subject and ROI are due to chance. However, we can analyze the percent of ROIs for a given subject that prefer the VDGLM to assess whether the amount of subject-level preference is due to chance. We compute whether the subject-level VDGLM preference is greater in the actual data compared to the simulated data. This test shows whether VDGLM preference is caused by overfitting to autocorrelation or whether there is true variance-related signal in the data. 

Model fitting and model comparison for 875 subjects took approximately 1/2 a day to fit on the UCI High Performance Cluster using in-house MATLAB code that can be found online \cite{vdglm_github}.



\subsection{ Results }


\begin{figure*}[!h]
     \centering
 		 \includegraphics[width=1\textwidth]{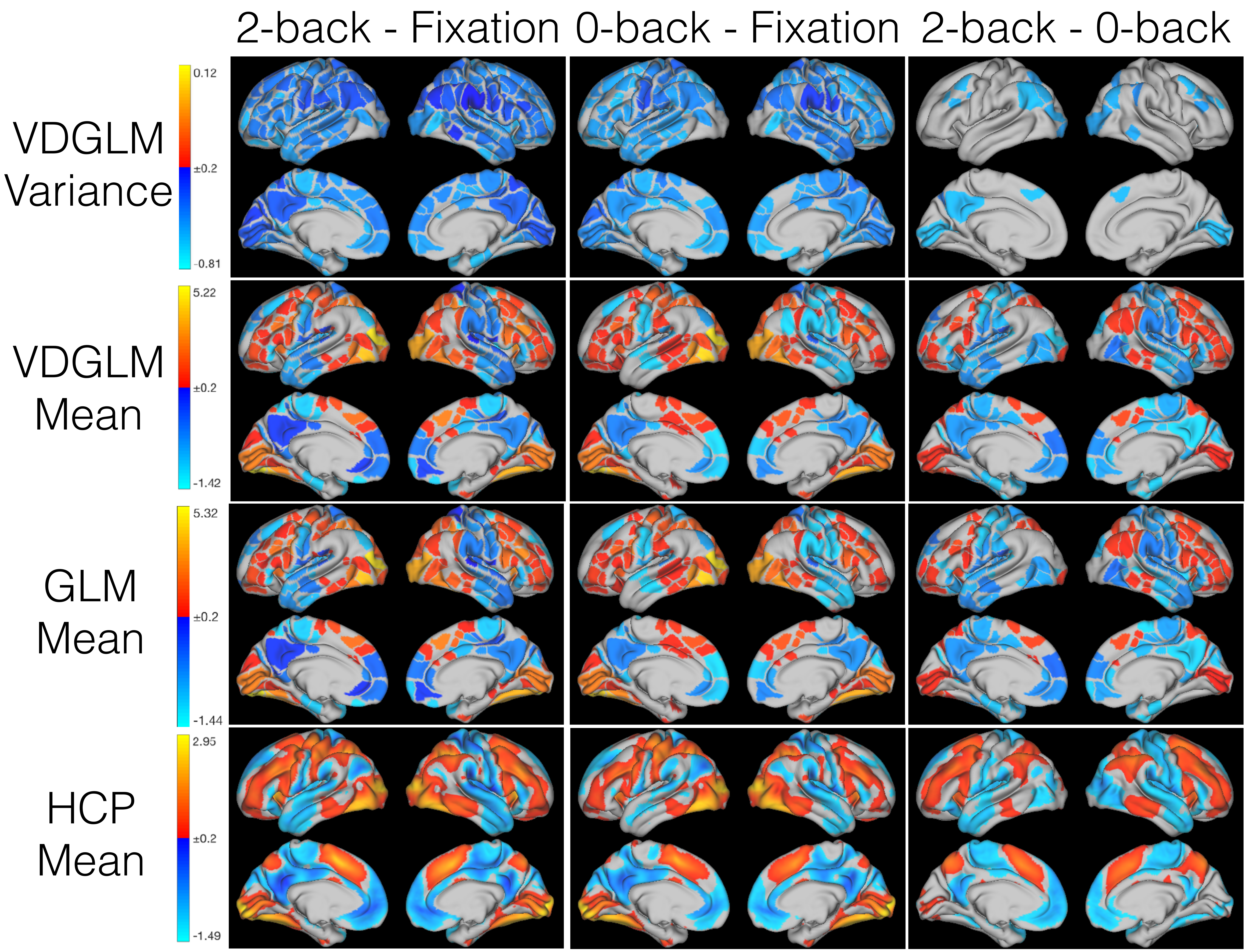}
		 \caption{ Group-wise Cohen's \textit{d} for the 2-back minus Fixation, 0-back minus Fixation, and 2-back minus 0-back contrasts. The top row shows VDGLM variance effects, the middle row shows VDGLM mean effects, and the bottom row shows voxel-wise GLM results from previous analysis. Maps are thresholded at (-0.2, 0.2). 
		 }
  \label{fig:vdglm_varcohensd_whs26}
\end{figure*}

The analyses we ran on the VDGLM were designed with three goals in mind. First, we wanted to test for the existence of effects on BOLD variation during working memory engagement. We did this by computing effect sizes of variance parameter estimates. Second, we wanted to see whether these effects occur in regions that are spatially orthogonal to regions that exhibit mean effects. To do this, we visually examined whether mean and variance effect sizes are correlated and plotted whole-brain visualizations of where mean and variance effects occur. Finally, we wanted to verify that the VDGLM provides a better account of the data than the GLM using model comparison metrics based on out-of-sample log-likelihood.

\subsubsection{ Group Level Effect Sizes }

The first goal in our analysis was to test for existence of variance effects caused by working memory engagement. We measured effects by computing Cohen's d over subjects. For each parameter contrast (2-back minus Fixation, 0-back minus Fixation, and 2-back minus 0-back) we plotted whole-brain Cohen's d (see Figure \ref{fig:vdglm_varcohensd_whs26}, bottom row). We also plotted Cohen's d for mean parameter contrasts (top row) to verify that the VDGLM preserves known mean effects. 

We found small and medium sized variance effects during the 2-back and 0-back tasks compared to fixation across much of the entire brain. Both 2-back and 0-back engagement evoked less BOLD variation (i.e., negative Cohen's d) compared to Fixation across the whole brain. For the 2-back minus 0-back contrast, variance Cohen's d was low for some areas in the default mode network, some areas of the dorsal attention network, some parts of visual cortex, and some parts of the fronto-parietal network. 

The VDGLM found mean effects that overlap existing HCP GLM results \cite{barch2013function} (see Figure \ref{fig:vdglm_varcohensd_whs26}). The 2-back minus Fixation and 0-back minus Fixation contrasts showed activation in the bilateral frontal-parietal network, bilateral visual cortex, and deactivation in the default mode network, including medial prefrontal cortex, posterior cingulate, and the occipital parietal junction. These same regions were activated,
but less intensely for the 2-back minus 0-back contrast.


In general, task engagement leads to both positive and negative mean effect sizes, but predominantly negative variance effect sizes.

\begin{figure*}[h!]
     \centering
  		 \includegraphics[width=1\textwidth]{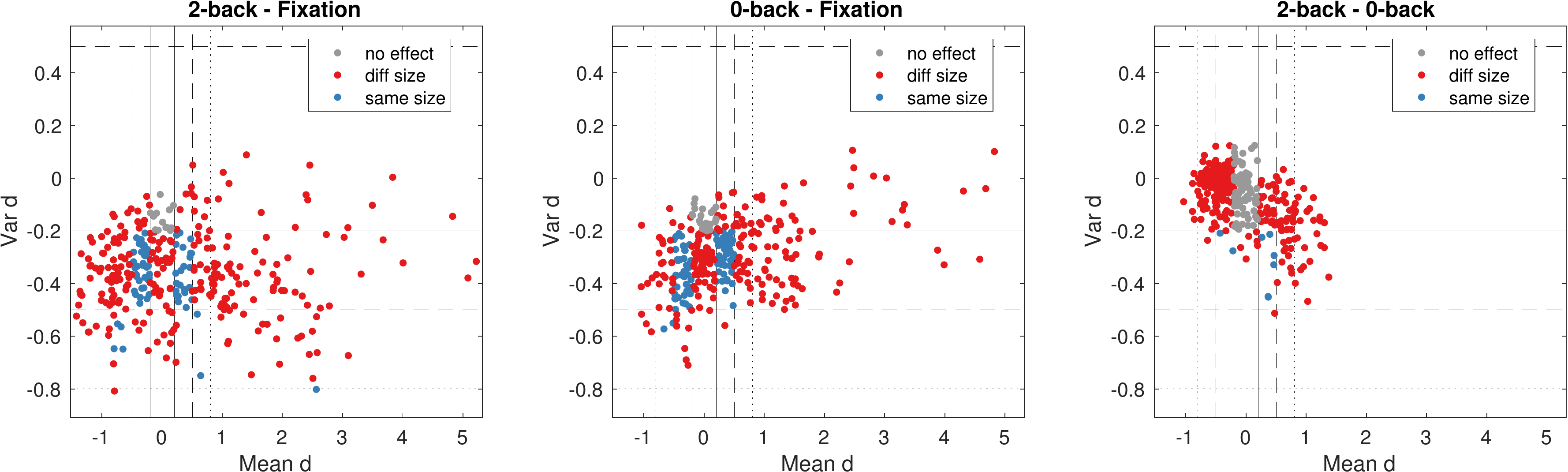}
		 \caption{The magnitude of mean and variance effects sizes for the 2-back minus Fixation, 0-back minus Fixation, and 2-back minus 0-back contrasts. Each circle represents an ROI. ROIs are grouped by whether they exhibit the same size effect (blue) in the mean and variance or different sized effects (red). The black lines indicate the small (solid), medium (dashed), and large (dotted) effect sizes. }
		 \label{fig::cohensscatter}
\end{figure*}

\subsubsection{ Orthogonality of Mean and Variance Effects } 

The second goal in our analysis was to examine whether the VDGLM finds variance effects that are orthogonal to known mean effects. To analyze the degree of orthogonality between mean versus variance effects, we plotted regional Cohen's d for the mean contrasts versus variance contrasts (See figure \ref{fig::cohensscatter}). We grouped ROIs by effect size. ROIs with the same effect size in the mean and variance are plotted in red, those with difference effect sizes in the mean and variance are plotted in blue, ROIs with no mean nor variance effects are plotted in gray. The small, medium, and large, effect thresholds are plotted by the solid, dashed, and dotted black lines, respectively. ROIs exhibited mean and variance effects that span all possible combinations of effect sizes, although there were no large variance effects for the 0-back minus Fixation nor 2-back minus 0-back contrasts (see Figure \ref{fig::cohensscatter}). In general, mean effects were larger than variance effects. Effects were also much larger for the 2-back minus Fixation and 0-back minus Fixation contrasts compared to the 2-back minus 0-back contrast. While there was slight negative correlation between mean Cohen's d and variance Cohen's d for the 2-back minus 0-back contrast ($R^2 = 0.31$), the other two contrasts were uncorrelated ($R^2 = -0.00294$ and $0.132$). Hence, mean and variance effects are orthogonal for the 2-back minus Fixation and 2-back minus 0-back contrasts.


\begin{figure*}[h!]
     \centering
 		 \includegraphics[width=1\textwidth]{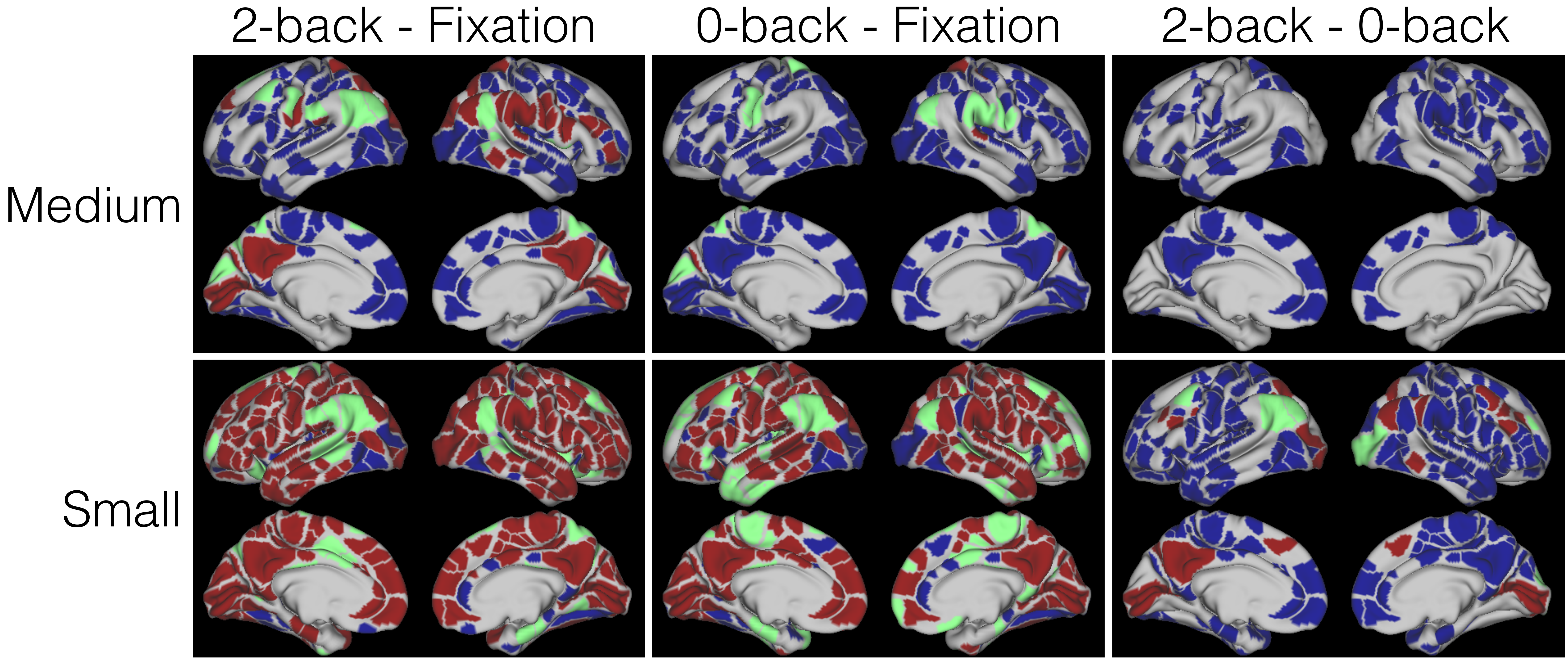}
		 \caption{ The figure shows which types of effects occur in which regions. Regions can have only a mean effect (blue), only a variance effect (green), or both effects (red). We plot effects for the 2-back minus Fixation, 0-back minus Fixation, and 2-back minus 0-back contrasts. We plot small (Cohen's d $\in \; [-0.2, 0.2]$) and medium (Cohen's d $\in \; [-0.5, 0.5]$) effects. }
  \label{fig:all_effects_whs26}
\end{figure*}

\subsubsection{ Spatial Distribution of Effects } 

Given that mean and variance effects were orthogonal for the 2-back minus Fixation and 2-back minus 0-back tasks, we wanted to see where each type of effect occurs in the brain. We plotted the type of small and medium effects that occur in each region (see Figure \ref{fig:all_effects_whs26}). A region with mean effect only is plotted in blue, variance effect only in green, and both effects in red. For the 2-back minus Fixation contrast, there are regions that exhibited all types of effects. Variance only effects occurred primarily in the default mode network, but also in sensorimotor mouth regions, and regions in the visual, dorsolateral attention, and cingulo-opercular networks. Mean only effects occurred in the frontoparietal, the auditory, cingulo-opercular, visual, and dorsolateral attention networks. Effects overlapped in some regions of the default mode, frontoparietal, visual, dorsal attention, and cingulo-opercular networks and in some sensorimotor regions. For the 0-back minus Fixation contrast, less regions exhibited medium variance effects than the 2-back minus Fixation contrast, suggesting that cognitive demand plays a role in the size of variance effects. There were, however, effects in sensorimotor areas used in control of the mouth and hand, a region in visual cortex, and some regions in the dorsolateral attention network. These effects were all present in the 2-back minus Fixation contrast as well. Mean 0-back minus Fixation effects also largely mirror the mean effects in the 2-back minus Fixation tasks. There were no medium sized variance 2-back minus 0-back contrasts. This suggests that while there were different sized effects between 2-back minus Fixation and 0-back minus Fixation, these differences in effect sizes were small. The main differences in variance effects tended to be caused by task engagement rather than cognitive load. 

\subsubsection{ Model Comparison }

The VDGLM inferred that engagement in a working memory task leads to less BOLD variation. In this section, we pursue a different question. Does a model with these additional parameters give a significantly better account of the BOLD time series than a model without them? To test this, we perform model comparisons between the VDGLM and the GLM. Since the VDGLM has more parameters and will trivially better fit the data, we use out-of-sample log-likelihood to check the VDGLM's ability to describe new unseen data. We perform a model comparison for each subject and ROI time series in a mass univariate approach. 

\begin{figure}[!h]
     \centering
  		 \includegraphics[width=0.4\textwidth]{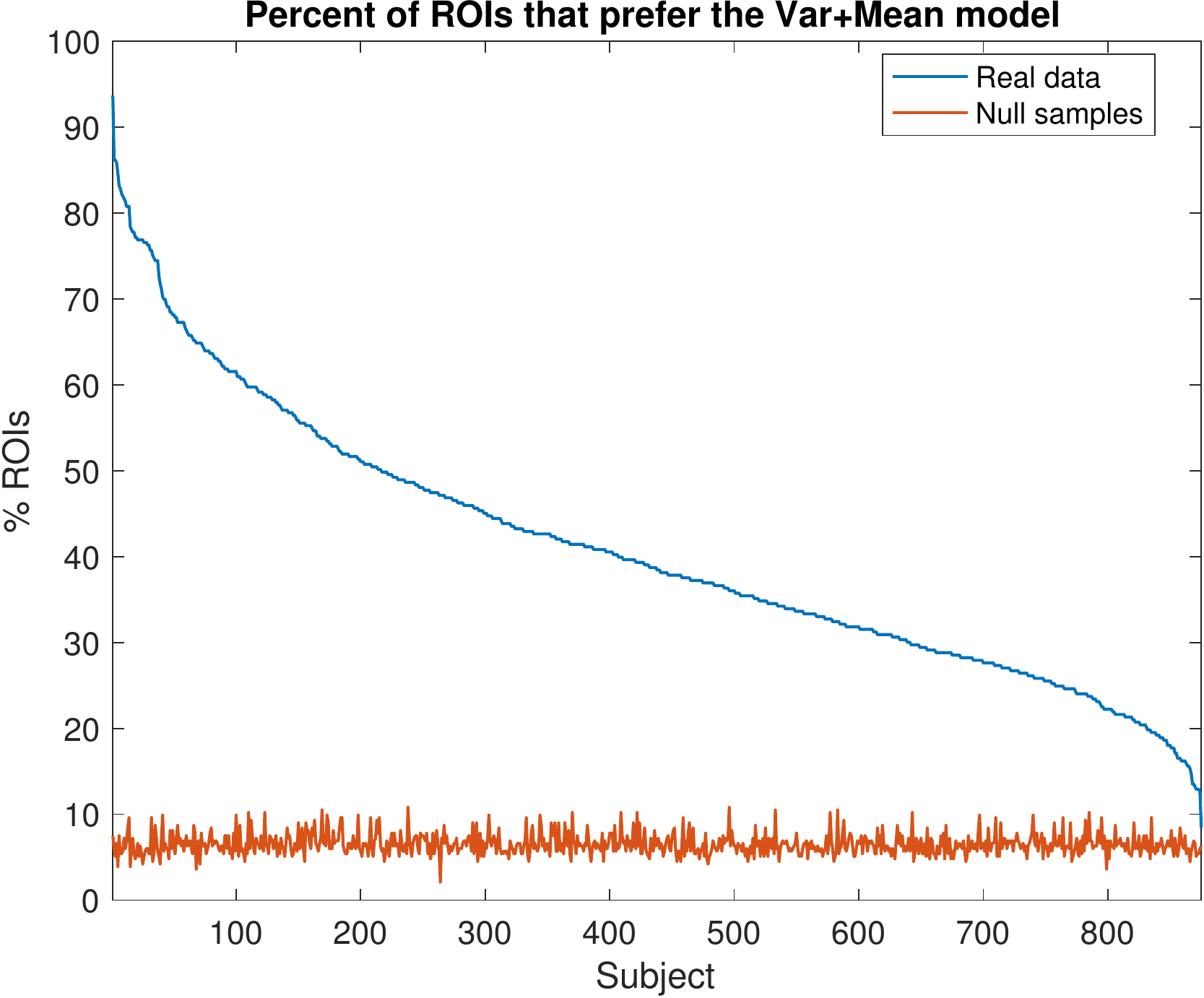}
		 \caption{ The percent of regions for each subject for which the VDGLM model better describes the HCP data (blue) and data simulated from the mean model (red). Subjects are ordered by percent of regions for which the VDGLM model has higher OOSLL. }
		 \label{preferred_whsim26}
\end{figure}

We found significant preference for the VDGLM model; 41\% of subjects/ROIs had higher OOSLL for the VDGLM model over the GLM in the real data (7\% in the simulated data). Model preference varies by subject and region (see Figure \ref{preferred_whsim26}). The figure plots the percentage of ROIs that prefer each model in the real data (blue) and the simulated data (red). The figure is ordered by a subject's proportion of ROIs that prefer the VDGLM model in the real data. A subject's percentage of regions that favored the VDGLM ranged from 14\% for GLM-leaning subjects to 73\% for VDGLM-leaning subjects. For all 875 subjects, some number of regions (but not all) preferred the VDGLM model. Similarly, for all 333 ROIs, some number of subjects (but not all) preferred the VDGLM model. 

To check that the VDGLM was not just fitting autocorrelation, we performed model comparisons for models fitted to data generated from the GLM with added autocorrelation. For the simulated data, only 7\% of subjects/ROIs preferred the VDGLM model. For every subject, the percent of ROIs that preferred the VDGLM model was larger in the real data than in the simulated data, indicating a significant preference for the VDGLM that was not just due to fitting to autocorrelation. 

\subsection{ Application Summary }

The working memory application highlights the usage of the VDGLM. We used the model to show that working memory task engagement was related to a decrease in variance compared to fixation. Variance effects and mean effects were not spatially correlated, suggesting that the VDGLM reveals distinct brain patterns not captured by mean-based approaches. We also found that the spatial distribution of mean effects was similar to the spatial distribution of voxel-wise mean effects from previous analyses \cite{barch2013function}. Variance tended to be reduced across the whole brain compared to baseline. We want to highlight that while many of the variance effect sizes in this application were fairly small, this will not necessarily be true in future applications. Importantly, if there exists some quantity of interest that consistently relates to small variance effect, then these small effects are worth studying. This is especially true in disease studies where discoveries have the potential to impact human lives \cite{rombouts2005altered,d2003alterations}. 

The application focused on testing for effects of BV, so we designed our application to reduce potential confounds. By using HCP data we minimized the effects of noise from CSF, large veins, and white matter; high resolution data collection (2mm) and registration to the cortical surface leads to less voxel-by-voxel overlap with these noise sources than compared to other data sources \cite{glasser2013minimal}. We corrected for head motion by including nuisance motion regressors and scrubbing particularly noisy volumes. We did not account for heart beat nor respiration, which are known to affect resting state BOLD variability \cite{biswal1996reduction, kannurpatti2008detection, kannurpatti2010neural, kannurpatti2011increasing}. However, neither source of noise could account for the variance effects we demonstrated. Since neither heart rate nor respiration are correlated with the task design, presence of these sources of noise increases variance during task. Thus we suspect that correcting for physiological noise in future studies would lead to larger variance effect sizes. We performed several post-hoc analyses to check that VDGLM preference was not related to to mean frame displacement, nor grand mean intensity scaling factor \cite{turner2015one} (adj$R^2 = -0.00114, -0.0011$, respectively). 

In this application, we fit a single VDGLM model and a single corresponding GLM model. In practice, we could fit several VDGLM models to test hypotheses of the form: ``should condition $C$ be be included in our model and does it affect the mean or the variance in BOLD activation''. In this set-up, each model takes the form of eq. \ref{VDGLM} and the conditions to be tested are defined by the entries of the mean and variance design matrices. For example, we could test a model with only intercept effects versus a model that allows each condition to affect the variance, but not the mean. Then model comparison indicates which experimental conditions are necessary in the model and whether those conditions are necessary as mean or variance regressors. This formulation allows us to define several nested models in the classical sense--i.e., that the set of mean regressors in the nested model is a subset of the regressors of the full model-- or in a novel way where a combination mean and variance regressors are nested. 

To effectively develop and test the VDGLM, we chose an ROI approach to have more reliable BOLD signal and a lower computational load. Application of the VDGLM to voxel-wise analyses is left to future work.

\section{ Discussion }

Traditional fMRI analyses treat BOLD variation as a `nuisance parameter' despite results linking BOLD variation to age, behavioral performance, and task engagement. The VDGLM fills this gap by providing a flexible framework for linking variance effects to experimental design. By directly incorporating the design matrix, the VDGLM can assess the independent contributions to BOLD variance from multiple experimental conditions while controlling for confounding factors. The VDGLM also controls for confounding between mean and variance effects; since both effects are modeled simultaneously, we can make inferences about one effect while controlling for the other. The VDGLM is fit in a mass univariate approach, which allows analysis at a more fine-grained resolution than previous empirical studies that analyzed latent structures of large spatial patterns in BV \cite{garrett2010blood, garrett2011importance, garrett2013moment}. Under the VDGLM framework hypothesis generation and comparison is easy; each hypothesis corresponds to an instantiation of the model and can be tested using model comparison. 

In our application, we showed that the VDGLM can be used to find variance effects caused by working memory engagement (Figure \ref{fig:vdglm_varcohensd_whs26}). We showed that these effects are spatially orthogonal to mean effects (Figures \ref{fig::cohensscatter}, \ref{fig:all_effects_whs26}) and finally, we compared the GLM and VDGLM and showed that VDGLM provides a better description of the data even while accounting for model complexity (Figure \ref{preferred_whsim26}). 

An important feature the VDGLM is the facility for modeling mean and variance simultaneously while allowing for orthogonal spatial inferences. In the BV-age fMRI literature, variance effects were orthogonal to mean effects \cite{garrett2010blood, garrett2011importance}. This trend generalized to our working memory application, where task engagement resulted in predominantly negative BV effects across the brain, but a mix of positive and negative mean effects. These results constitute a growing body of evidence that BV is a novel dimension for studying brain function. 

We want to highlight that there are alternatives to the methodological choices we made in our application. Alternative choices can be made regarding 1) the prewhitening model, 2) the inference statistic or effect size estimate, 3) the model comparison metric and 4) the method for assessing model comparison significance (see Figure \ref{fmri_flow}). The 
prewhitening model (1) and comparison metric (3) can easily be substituted for another model and metric, and the additional model comparison significance test (4), while powerful, is not necessary in most standard analyses. Using an alternative choice of inference statistic (2) may require further work. Our choice to use effects size was motivated by the use of effect size in previous analyses \cite{van2013wu} and the ease of using a statistic that depends solely on parameter estimates. In an effort to provide alternative inference statistics, we developed approximate t-tests for variance effects. However, we found that the estimates were sensitive to the condition number of the Hessian matrix specified by the VDGLM (which is required for computation of approximate t-tests). 
We tested the accuracy of the approximate t-tests for mean effects by comparing them to standard t-tests made by the GLM. While we found that while they were close for most subjects and ROIs, for other subjects with poorly conditioned Hessian matrices the t-tests tended to be unrealistically large. While any individual data point could be excluded from group-level analysis using condition number threshold, we found this approach too cumbersome for a framework aimed at general public use. Development of well-behaved statistics for inference is ongoing work.

Many different measures of BV have been used in past fMRI studies: empirical variance \cite{he2011scale}, parametric variance \cite{wutte2011physiological}, block-normalized standard deviation \cite{garrett2010blood, garrett2013modulation}, and mean squared successive difference (MSSD) \cite{leo2012increased, samanez2010variability}. The goals of mean squared successive difference and block-normalized standard deviation are to measure the variance not accounted for by mean trends in the data. Since the VDGLM models the variance/standard deviation in BOLD activation after accounting for the mean trend, its variance parameters can be conceptualized to measure a construct similar to mean squared successive difference or block-normalized variance (but where blocks are convolved with the canonical HRF). The parametrized model in Wutte et al. 2011 is similar to the VDGLM, but uses a mixing parameter to capture shared variance between task and fixation blocks rather than modeling the variance as a function of convolved experimental design. We expect that this approach leads to similar results, but with the caveat that it only incorporates a single experimental condition. Lastly, we consider the inter-quartile range, which is not used in the VDGLM and to our knowledge has not been used in fMRI analysis to date. The goal of the inter-quartile-range is to summarize the dispersion while limiting the effects of any highly outlying time points. In our application, we used scrubbing to a similar effect by manually removing any outlying time points and recommend this approach if large outliers are present. 

The VDGLM could be improved by implementing it in a Bayesian framework. Bayesian frameworks would allow us to make more robust inferences, incorporate prior beliefs about regions likely (or unlikely) to exhibit BV effects, and to better quantify model comparisons. The main disadvantage of Bayesian methods, and the reason we did not develop a Bayesian VDGLM, is the computational complexity of inference. 


The VDGLM could also be improved by transforming the variance so that we did not need to enforce positivity. Log transformations have been widely used for covariance and variance estimation \cite{pourahmadi2011covariance}, however in the case of the VDGLM lead to a drastic conceptual change in the model. Since the VDGLM incorporates the design matrix into its variance formulation, the exponential transformation results in a variance parameters that are raised to the power of elements of the design matrix. We opted to keep the VDGLM as an additive variance model that requires constraints rather than as a multiplicative variance model so that parameters were more interpretable. However, we expect that the study of variance transformations could lead to stronger inferences in future work. 

\section{ Conclusion } 

Studies have demonstrably shown that variance in BOLD activation is a functional construct orthogonal to mean BOLD that should be taken into account in future imaging analyses.

This work developed the VDGLM, a coherent statistical framework for incorporating BV into standard fMRI analyses. The VDGLM was motivated by strong evidence that variance in BOLD activation is linked to individuals and behavior. The VDGLM can be easily applied in any experimental setting and will allow for increased ease and flexibility in research on BOLD variability. We expect that it will lead to exciting new discoveries relating BOLD variability to human characteristics and behavior. 

\section{ Acknowledgments } 
This work was supported by a National Sciences Foundation Integrative Strategies for Understanding Neural and Cognitive Systems Collaborative Research Grant (1533500 and 1533661).

The authors have no conflicts of interest to declare.

\begin{appendices}

\section{ VDGLM Optimization}\label{optimization}


We perform maximum likelihood estimation using Trust-region optimization (TRO). TRO is an iterative procedure for minimization. At each iteration, TRO locally approximates the negative log likelihood function using a Taylor expansion and finds a minimum within that step's trust-region, i.e. the region for which the local approximation accurately approximates the objective function. We used a built-in function in MATLAB that restricts the local approximation to a 2-dimensional subspace to allow for faster convergence. The algorithm locally minimizes along the two-dimensional subspace spanned by the direction of steepest descent and one of either a) the approximate newton direction, if it exists, or b) the direction of negative curvature \cite{byrd1988approximate}. For a time series $y$ from a single subject and region, we minimize the negative log likelihood $ - \log p(y | \theta) $ where the likelihood is defined: 

\small 
\begin{equation*} 
 p(y, \theta) = \Big [ \prod_{t=1}^{T} p(y_t | \beta,v, X^M_t, X^V_t) \Big ] 
 \end{equation*} 
 \normalsize 
 
\noindent where $X_t^M$ is a $[1 \times p_M]$ vector: the single row of the mean design matrix at time $t$. $X_t^V$ is a $[1 \times p_V]$ vector: the single row of the variance design matrix at time $t$. For single point in the time series, the likelihood is

\small 
  \begin{align*} 
 p(y_t | &\beta, v, X^M_t, X^V_t) = \\ 
 &\frac{1}{\sqrt{2\pi X^V_t v}} \exp \Big ( \frac{-1}{2X^V_t v} (y_t - X^M_t \beta )^2 \Big ). 
 \end{align*} 
 
 \normalsize 
 
 \noindent and we can compute the joint log-likelihood as the product of the log-likelihoods for each point: 
 
 
 \small 
 \begin{align*} 
 \log p( y | & (\theta ) = \\
 &-\frac{T}{2}\log2\pi + \\
 &\Big [ \sum_{t}^{T}{ -\frac{1}{2}\log (X^V_t v) - \frac{1}{2X^V_t v} (y_t - (X^M_t \beta ) )^2 } \Big ]
 \end{align*} 

\normalsize

\noindent subject to the inequality constraint that the variance is nonzero, i.e.: 

\small 
\begin{equation*}
  X^V_tv > 0 \; \forall \; t \in \{1, ..., T\} 
\end{equation*}
\normalsize 

\noindent This constraint can be conceptualized in a Bayesian setting as a uniform prior over the constrained area. Each iteration of the trust-region algorithm uses a Newton-Raphson step to update. We supply the analytical gradients: 

\small 
\begin{equation*}
\frac{ \partial \log p(\beta | \bullet )}{\partial \beta} =(X^M_t)^T \Big [ \sum_{t=1}^T{ \frac{ (y_t - X^M_t \beta)}{ X^V_t v } } \Big ] 
\end{equation*}

\begin{equation*} 
\frac{\partial \log p(v | \bullet) }{ \partial v} = (X^V_t)^T \Big [ \sum_{t=1}^T{ \frac{-1}{2X^V_t v} + \frac{(y_t - X^M_t \beta )^2}{2(X^V_t v)^2}} \Big ].
\end{equation*} 

\normalsize 

We stop the optimization routine when the magnitude of the gradient is smaller than 1e-6, the change in objective value is smaller than 1e-6, the size of the trust region is below 1e-6, or the optimization routine reaches 1000 iterations.

\end{appendices}

\bibliographystyle{apalike}
\bibliography{advancement2} 

\end{document}